\documentclass[sigconf]{acmart}
\usepackage[utf8]{inputenc}
\usepackage{newunicodechar}
\newunicodechar{́}{\'{} }
\usepackage{fancyvrb}
\usepackage{listings}
\usepackage{xcolor}
\usepackage{caption}
\usepackage{booktabs} 
\usepackage{csquotes}
\usepackage{graphicx}
\usepackage[most]{tcolorbox}
\tcbset{colback=gray!5!white, colframe=gray!80!black, boxrule=0.5pt, arc=2pt, left=6pt, right=6pt, top=6pt, bottom=6pt}
\usepackage{array}
\usepackage{placeins}
\usepackage{float}
\usepackage{longtable}
\usepackage{tabularx}
\tcbset{
  resultbox/.style={
    colback=black!5!white,
    colframe=black!75!black,
    title font=\bfseries,
    coltitle=white,
    boxrule=0.5mm,
    arc=2mm,
    left=2mm,
    right=2mm,
    top=4mm,
    bottom=0.5mm
  }
}
\newcommand{\patterntitle}[1]{\textbf{\textit{#1}:}\ignorespaces}
\AtBeginDocument{%
  }

\setcopyright{acmlicensed}
\copyrightyear{2018}
\acmYear{2018}

\acmConference[SBES '25]{Brazilian Symposium on Software Engineering}{September 22--26, 2025}{Recife, PE}
\setcopyright{none}
\renewcommand\footnotetextcopyrightpermission[1]{}

\author{Arthur Bueno}
\affiliation{
  \institution{FACOM/UFMS}
  \state{MS}
  \country{Brazil}}
\email{arthur.ramires@ufms.br}

\author{Bruno B. P. Cafeo}
\affiliation{%
  \institution{UNICAMP}
  \state{SP}
  \country{Brazil}}
\email{cafeo@ic.unicamp.br}

\author{Maria Istela Cagnin}
\affiliation{%
  \institution{FACOM/UFMS}
  \state{MS}
  \country{Brazil}}
\email{istela.machado@ufms.br}

\author{Awdren Fontão}
\affiliation{%
  \institution{FACOM/UFMS}
  \state{MS}
  \country{Brazil}}
\email{awdren.fontao@ufms.br}

\renewcommand{\shortauthors}{Bueno et al.}

\begin{document}

\title[Socio-Technical Smell Dynamics in Code Samples]{Socio-Technical Smell Dynamics in Code Samples: A Multivocal Review of Their Emergence, Evolution, and Co-Occurrence}

\begin{abstract}
\textbf{Context and Motivation:} Code samples play an important role in open-source ecosystems (OSSECO), serving as lightweight artifacts that support knowledge transfer, onboarding, and framework adoption. Despite their instructional relevance, these samples are often governed informally, with minimal review and unclear ownership, which increases their exposure to socio-technical degradation. In this context, the co-occurrence and longitudinal interplay of \textit{code smells} and \textit{community smells} become particularly relevant. While each type of smell has been studied in isolation, little is known about how community-level dysfunctions anticipate or exacerbate technical anomalies in code samples over time. \textbf{Objective:} This study investigates how code and community smells emerge, co-occur, and evolve within code samples maintained in OSSECOs. \textbf{Method:} A Multivocal Literature Review protocol was applied, encompassing 18 peer-reviewed papers and 17 practitioner-oriented sources (2013–2024). Thematic synthesis was conducted to identify recurring socio-technical patterns. \textbf{Results:} Nine patterns were identified, showing that community smells often precede or reinforce technical degradation in code samples. Symptoms such as "radio silence" and centralized ownership were frequently associated with persistent structural anomalies. Additionally, limited onboarding, the absence of continuous refactoring, and informal collaboration emerged as recurring conditions for smell accumulation. \textbf{Conclusion:} In OSSECOs, particularly within code samples, community-level dysfunctions not only correlate with but often signal maintainability decay. These findings underscore the need for socio-technical quality indicators and lightweight governance mechanisms tailored to shared instructional artifacts.
\end{abstract}


\maketitle
\section{Introduction}

Code samples are important artifacts in software ecosystems, particularly in open-source environments (OSSECOs), where they are used to illustrate framework usage, propagate best practices, and support rapid onboarding of new contributors~\cite{Menezes2019, Menezes2022, Melo2023, fontaoDevRel}. These samples are typically small-scale, self-contained modules designed to be readable and instructional. However, despite their instructional importance, they are frequently treated as second-class citizens, often excluded from automated testing pipelines, static analysis tools, and formal review processes.

This treatment makes code samples especially vulnerable to socio-technical degradation \cite{Menezes2019}. Contributions frequently originate from external developers within the broader SECO—many of whom are unfamiliar with project-specific conventions or architectural constraints \cite{Melo2023}. In the absence of shared norms, clearly defined roles, or structured maintenance routines, these examples may accumulate low-quality code and undocumented practices \cite{Subramanian2014}. For instance, if a developer encounters a sample exhibiting a \textit{Large Class} or missing modular structure, the lack of clarity may lead to incorrect reuse, onboarding friction, or even abandonment of a contribution attempt~\cite{Nasehi2012}.

Technical and social degradation in such contexts can be described through the lenses of \textit{code smells} and \textit{community smells}. Code smells refer to structural symptoms that compromise maintainability, such as \textit{Large Class}, \textit{Long Method}, or \textit{Feature Envy}~\cite{fowler2018refactoring, ARCELLI_FONTANA2013}. Community smells, in contrast, reflect patterns of collaboration breakdown, including \textit{Lone Wolf} contributors, \textit{Organizational Silo}, and \textit{Radio Silence}~\cite{TAMBURRI2021, Almarimi2020, Almarimi2022}. These social anomalies have been associated with delayed maintenance, reduced transparency, and hindered knowledge transfer~\cite{DeStefano2021}.

While these smells have been extensively studied in isolation, a growing body of evidence suggests they are interrelated, particularly in informally maintained artifacts like code samples. Socioorganizational dysfunctions often act as precursors to technical anomalies~\cite{Palomba2018, Mumtaz2022}, and these interdependencies are rarely captured by traditional quality assurance mechanisms. For example, in environments marked by contributor turnover or limited peer feedback, smells may persist across releases unnoticed.

Despite this potential interplay, the co-occurrence and reinforcement of code and community smells in code samples remain poorly understood. Existing studies tend to focus on core modules or monolithic systems, overlooking lightweight artifacts that play key roles in learning and adoption. Moreover, little is known about how smells evolve longitudinally or how their interactions shape sustainability and collaboration at the ecosystem level.

To address these gaps, this study applies a \textit{Multivocal Literature Review} (MLR)~\cite{GAROUSI2019101}, synthesizing findings from peer-reviewed literature and grey sources (e.g., blog posts, practitioner reports, community forums). The review focuses explicitly on the socio-technical dynamics of smells in open-source code samples and aims to uncover patterns that explain their emergence, evolution, and co-occurrence.

This investigation is guided by the following research questions:
\begin{itemize}
    \item \textbf{RQ1:} What socio-technical conditions and practices lead to the emergence of Code Smells and Community Smells in code samples?
    \item \textbf{RQ2:} How do Code Smells and Community Smells evolve over time in open-source code samples, and how can this evolution be detected?
    \item \textbf{RQ3:} How do Community Smells signal or reinforce the emergence and persistence of Code Smells in code samples?
\end{itemize}
By analyzing these aspects, the review provides empirical foundations for the development of socio-technical quality indicators, informs preventive practices, and advances the understanding of how code samples in OSSECOs degrade in the absence of formal governance.

\section{Background and Related Work}

\subsection{Code Smells and Community Smells}

Code smells and community smells are widely acknowledged in the software engineering literature as indicators of quality degradation in software systems \cite{S16}. \textit{Code smells} denote suboptimal design choices or structural issues within the source code that, while not inherently faulty, may negatively affect maintainability, extensibility, or comprehension over time~\cite{fowler2018refactoring}. Common examples include \textit{Long Method}, \textit{Large Class}, and \textit{Feature Envy}~\cite{ARCELLI_FONTANA2013}, which are often addressed through refactoring practices. However, the detection of these smells---typically performed using automated tools---tends to occur without contextual consideration of the development environment or the collaborative dynamics surrounding the artifact. As a result, smell remediation is neither systematic nor guaranteed, particularly in artifacts that are loosely maintained or excluded from structured quality assurance workflows.

In contrast, \textit{community smells} reflect socio-organizational dysfunctions within development teams, often manifesting through patterns such as \textit{Lone Wolf}, \textit{Organizational Silo}, and \textit{Radio Silence}~\cite{TAMBURRI2021, Almarimi2022}. These smells are typically inferred from social and contribution metadata, and have been associated with coordination breakdowns, centralized expertise, and contributor disengagement~\cite{DeStefano2021}. Systematic reviews and recent empirical studies~\cite{Palomba2023, Porta2024, Chen2024} have emphasized their negative impact on collaborative efficiency, knowledge transfer, and software sustainability.

Despite the growing body of work on both categories of smells, their joint manifestation and evolution in \textit{code samples}—lightweight, instructional artifacts—remain largely underexplored \cite{Melo2023}. Unlike core modules, code samples are frequently developed with minimal governance, lacking defined ownership, review practices, or integration into continuous quality pipelines. Such informal conditions make them especially susceptible to the emergence and persistence of both code and community smells~\cite{Mumtaz2022, Baabad2020}. Investigating the co-occurrence and temporal interplay of these smells in code samples, as pursued in RQ2 and RQ3, is therefore crucial to understanding how socio-technical degradation unfolds in this class of artifacts.

\subsection{Code Samples in Open-source Software Ecosystems (OSSECOs)}

Code samples are lightweight, self-contained artifacts meant to illustrate usage patterns, support onboarding, and promote framework adoption within OSSECOs~\cite{Menezes2019, Menezes2022, Melo2023}. \textcolor{black}{Our definition of code samples \cite{8870139, MENEZES2022111146} characterizes them as complete, executable projects in official repositories, distinguishing them from isolated snippets. In this context, maintainability is a developer's ability to understand, execute, and adapt samples for learning. The observed degradation, which includes structural complexity, outdated dependencies, and misaligned configurations, directly impacts this. Our analysis isolates this degradation within code samples as distinct instructional artifacts.} Their role is especially relevant in open-source contexts, where they serve as first points of contact for new contributors. However, due to their instructional nature and auxiliary status, samples are frequently excluded from quality assurance pipelines and formal review processes~\cite{Subramanian2014, Yamashita2013, Annunziata2023}.

Unlike traditional systems that benefit from defined ownership and structured reviews, code samples are often maintained informally by external or occasional contributors. This leads to fragmented maintenance, high turnover, and low visibility in governance mechanisms~\cite{Melo2023, MoradiJamei2021}. Empirical studies and practitioner accounts have shown that smells introduced in these artifacts are rarely detected or corrected, especially when the surrounding team lacks process uniformity~\cite{Almarimi2022, Lambiase2024}.

Such conditions directly relate to RQ1 and RQ3: they reveal how structural and social anomalies originate and persist in loosely governed modules. A single contributor acting as a \textit{Lone Wolf} may introduce unresolved smells, while \textit{Radio Silence} may allow quality regressions to propagate. Understanding these dynamics in code samples is important for capturing early signals of decay within OSSECOs.

\subsection{Co-occurrence and Evolution: Gaps in Literature}

Although the evolution of smells has been studied independently, the intersection between code and community smells remains insufficiently addressed. One study found that social fragmentation often precedes the accumulation of persistent structural smells, suggesting a reinforcing effect between social and technical debt~\cite{Palomba2018}. Additional findings show that organizational misalignments may not only delay remediation, but also shape how smells spread across versions~\cite{Mumtaz2022, Tuna2024}.

Still, the causal mechanisms remain unclear. Some authors observe that process smells, like informal review routines, do not always result in structural anomalies~\cite{Tuna2024}, while others emphasize that the absence of onboarding pipelines contributes to sustained technical flaws~\cite{Wang2023}. Tools and datasets capable of tracking longitudinal evolution—such as those proposed in~\cite{Almarimi2023, Mahbub2024}—rarely focus on non-core artifacts like code samples, leaving blind spots in monitoring practices.

Recent contributions call for integrated analyses that combine contributor behavior with smell propagation models~\cite{Chen2024, Porta2024}. These gaps motivate RQ2 and RQ3 in this study: how do community smells affect the lifecycle of code smells, and how can such entangled phenomena be captured in shared instructional artifacts?

\subsection{Motivation for a Multivocal Perspective}

This review adopts a Multivocal Literature Review (MLR) approach~\cite{GAROUSI2019101, Soldani2021grey}. Grey sources—such as blog posts, technical forums, and community documentation—offer rich narratives on informal practices, role ambiguity, and tool skepticism that shape how smells manifest and are (not) addressed in practice.

Prior MLRs in software engineering have demonstrated that grey sources can provide early evidence of practical challenges that are only later formalized in academic research~\cite{Soldani2021grey, GAROUSI2019101}. This is particularly true in OSSECOs, where community-maintained artifacts are governed through hybrid (often undocumented) processes~\cite{fontaoMSECOCERT}.

Therefore, this study complements peer-reviewed evidence with practitioner perspectives to build a comprehensive understanding of smell dynamics in code samples—particularly in light of underreported phenomena such as smell co-occurrence, informal resolution strategies, and socio-technical invisibility.

\section{Research Method}
\label{sec:research_method}

This study adopts a \textit{Multivocal Literature Review} (MLR) method to investigate how \textit{Code Smells} and \textit{Community Smells} emerge, co-occur, and evolve in open-source code samples. 

\textcolor{black}{To better articulate our methodological stance, our study adopted an aggregated, cross-ecosystem approach. While our inclusion criteria (IC2) ensured all studies were situated within OSSECOs, we deliberately did not focus on specific domains (e.g., Apache, Android). The objective was to identify generalizable socio-technical patterns that transcend any single ecosystem. This approach is supported by literature \cite{MANIKAS201684}, which highlights the value of aggregated studies for establishing a foundational understanding of recurring phenomena. Our work thus aims to build a generalizable theory of smell co-occurrence in the broader OSSECO landscape.}

\textcolor{black}{From broader studies, we extracted only segments directly addressing code samples. For instance, from a general study on technical debt, we analyzed only the parts concerning sample repositories. This clarification expands upon our interpretive process, ensuring all findings are grounded in the context of instructional artifacts as defined in IC6 of our MLR protocol (Table~\ref{tab:gqm_mapping}).}

MLRs integrate evidence from both formal academic studies and grey literature sources, providing a broader and more context-sensitive synthesis of socio-technical phenomena in Software Engineering~\cite{GAROUSI2019101, Soldani2021grey}.

The research method protocol is organized in two complementary phases:
\begin{itemize}
    \item An \textbf{formal literature mining phase}, focused on identifying peer-reviewed studies that address the causes, evolution, and interdependencies of code and community smells in the context of open-source development;
    \item A \textbf{grey literature phase}, aimed at extracting practitioner-reported insights from blog posts, community forums, white papers, and technical repositories related to the practical maintenance of code samples in software ecosystems.
\end{itemize}

To ensure conceptual alignment across the review process, the \textit{Goal-Question-Metric} (GQM) approach~\cite{basili1994gqm} was applied. This approach enables the explicit mapping between the overarching research goal, guiding questions, and the types of data to be extracted. Table~\ref{tab:gqm_mapping} presents the structured mapping used in this study.

\begin{table}[htbp]
\centering
\caption{Mapping of Goal, Research Questions, and Extracted Metrics (GQM)}
\label{tab:gqm_mapping}
\footnotesize
\begin{tabular}{|p{0.7cm}|p{7.0cm}|}
\hline
\multicolumn{1}{|c|}{\textbf{Goal}} & \multicolumn{1}{p{7.0cm}|}{Understand how Code Smells and Community Smells co-occur and evolve in open-source code samples.} \\
\hline
\multicolumn{2}{|c|}{\textbf{Research Questions and Corresponding Metrics}} \\
\hline
\textbf{RQ} & \textbf{Question and Extracted Metrics} \\
\hline
\textbf{RQ1} & \textbf{Question:} What socio-technical conditions and practices lead to the emergence of Code Smells and Community Smells in code samples? \newline
\textbf{Metrics:} Reported causes and triggers; collaboration models; review and maintenance practices; ownership and role structure; onboarding mechanisms. \\
\hline
\textbf{RQ2} & \textbf{Question:} How do Code Smells and Community Smells evolve over time in open-source code samples, and how can this evolution be detected? \newline
\textbf{Metrics:} Temporal analysis strategies; detection tools; persistence and decay indicators; contributor transitions; monitoring workflows. \\
\hline
\textbf{RQ3} & \textbf{Question:} How do Community Smells signal or reinforce the emergence and persistence of Code Smells in code samples? \newline
\textbf{Metrics:} Co-occurrence episodes; symptom cascades; contributor roles; unresolved issues; interaction breakdowns; reinforcement patterns. \\
\hline
\end{tabular}
\end{table}

Each phase of the MLR followed a structured and documented protocol. In the formal literature mining phase, the selection was based on a refined search string applied to IEEE Xplore, ACM Digital Library, and Scopus. The search strategy incorporated multiple thematic axes—code samples, technical debt, collaboration structures—combined using Boolean logic. A quasi-gold standard approach~\cite{GAROUSI2019101} was applied to validate coverage by checking whether a core set of highly relevant papers was retrieved.

The grey literature phase used effort-bounded Google searches~\cite{GAROUSI2019101}, targeting domains such as Dev.to\footnote{https://dev.to/}, Reddit\footnote{https://www.reddit.com/}, Medium, and Zenodo. \textcolor{black}{This strategy involved screening the first 10 pages of results for each query and stopping when two consecutive pages yielded no new relevant sources, indicating a point of theoretical saturation for the scope of this review.} Selection criteria emphasized accessibility, authorship traceability, and relevance to the guiding questions. A three-step filtering process—title/scope screening, full-text analysis, and thematic validation—was applied to select high-quality non-academic sources. 

\textcolor{black}{Evidence was mapped to RQs using the GQM framework (Table~\ref{tab:gqm_mapping}), which defines a distinct scope for each RQ. The first author conducted the synthesis via a structured, multi-phase approach. First, open coding was performed on the data using the GQM-aligned extraction form. The initial codes (e.g., “absence of peer review”) were then iteratively refined over several passes to ensure consistent application. Finally, the refined codes were systematically grouped into nine patterns using axial and selective coding. Each piece of evidence was categorized by its primary focus: causal conditions (RQ1), temporal evolution (RQ2), or reinforcing mechanisms (RQ3). For example, “developer responsibility” was mapped to RQ1 when described as a lack of initial ownership, but to RQ3 when described as sustaining unresolved issues. This operational approach grounded all allocation decisions in the textual evidence. The complete mapping is documented in our supplementary artifact~\ref{sec:data}.}

\section{Formal Literature Mining}
\label{sec:literature_mining}

\subsection{Rationale and Scope}

The formal literature mining phase aimed to identify empirical studies that explore the emergence, evolution, and interplay of \textit{Code Smells} and \textit{Community Smells}, particularly in the context of code samples within software ecosystems. While existing research frequently isolates technical or social anomalies~\cite{Palomba2018, TAMBURRI2020}, this review sought studies that examine their intersection or provide insight into socio-technical quality degradation over time.

\subsection{Control Studies and Validation}

To validate the coverage and focus of the search strategy, four control studies were selected based on their relevance to the research questions and recognition in the literature:

\begin{itemize}
    \item \textbf{C1}: \cite{Palomba2018} — Persistence of code smells and the role of communication breakdowns;
    \item \textbf{C2}: \cite{Almarimi2020} — Detection techniques for community smells in open-source ecosystems;
    \item \textbf{C3}: \cite{ARCELLI_FONTANA2013} — Temporal evolution of code smells in large-scale systems;
    \item \textbf{C4}: \cite{TAMBURRI2020} — Socio-technical impacts of community smells.
\end{itemize}

The final query string was refined iteratively until at least three of these four studies were retrieved by default, ensuring baseline adequacy.

\subsection{Search Strategy}

The search string was constructed through six refinement cycles, incorporating expert feedback. It targeted three thematic axes: technical debt in instructional artifacts, socio-technical collaboration, and open-source quality practices. The final string was:

\begin{center}
\texttt{(“code samples” OR “sample code”) \textbf{AND} (“code smells” OR “community smells” OR “social debt” OR “technical debt”) \textbf{AND} (“developer communication” OR “team structure” OR “developer communities” OR “open source projects”)}
\end{center}

This query was applied to IEEE Xplore, ACM Digital Library, and Scopus—chosen for their relevance to software engineering research and empirical studies.

\subsection{Inclusion and Exclusion Criteria}

\textbf{Inclusion Criteria (IC):}
\begin{itemize}
    \item IC1: Empirical studies addressing code and/or community smells;
    \item IC2: Studies analyzing open-source projects that belong to or operate within a software ecosystem (SECO);
    \item IC3: Use of metrics, automated tools, or temporal data (e.g., evolution over versions);
    \item IC4: Studies that provide evidence relevant to at least one of the guiding research questions, including those addressing evolution and co-occurrence dynamics;
    \item IC5: Publications written in English;
    \item IC6: Studies referring explicitly to code samples or equivalent learning artifacts (e.g., documentation-attached examples), excluding trivial code samples.
\end{itemize}

\textbf{Exclusion Criteria (EC):}
\begin{itemize}
    \item EC1: Theoretical, position, or review papers lacking empirical evidence;
    \item EC2: Papers without full-text access;
    \item EC3: Studies focusing on non-software domains or unrelated artifacts (e.g., education, robotics, hardware).
\end{itemize}

\subsection{Selection Process and Corpus Definition}
\label{subsec:selection_process}

The selection process was conducted in four stages (Figure 1), each applying the inclusion and exclusion criteria with increasing depth:
\begin{enumerate}
    \item \textbf{Initial Retrieval:} 63 documents were retrieved from the three databases;
    \item \textbf{Deduplication:} Duplicates across databases were identified by metadata matching and removed. The deduplication step focused on eliminating identical studies indexed differently across platforms. This yielded 37 unique entries for screening;
    \item \textbf{Title and Abstract Screening:} 28 studies were excluded based on topical misalignment (e.g., education, unrelated artifacts). Nine candidate studies remained;
    \item \textbf{Full-text Screening and Snowballing:} All nine studies were retained. Backward and forward snowballing identified nine additional studies, resulting in a final corpus of 18.
\end{enumerate}

\begin{figure}[h]
\centering
\includegraphics[width=\linewidth]{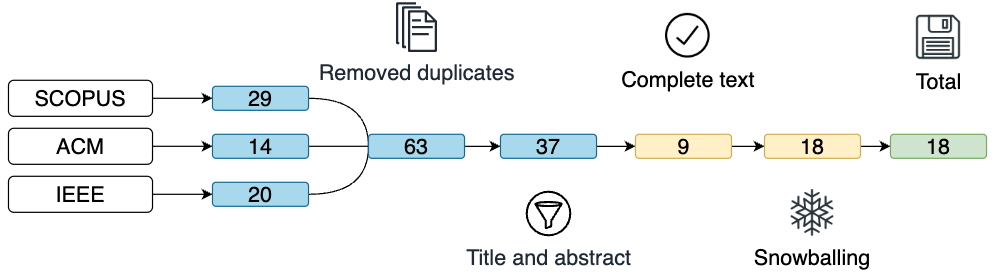}
\caption{Formal literature selection process}
\label{fig:slr_flow}
\end{figure}

This corpus offers a diverse but focused representation of how code and community smells are studied in open-source contexts, particularly regarding their emergence, evolution, and socio-technical interdependencies.

\subsection{Included Studies}
\label{subsec:slr_included}

The final corpus comprises 18 empirical studies (S1–S18) ~\cite{S2,S3,S4,S5,S6,S7,S8,S9,S10,S11,S12,S13,S14,S15,S16,S17,S18}, each selected through systematic screening and full-text analysis. The studies were coded thematically and mapped to the research questions according to their methodological and conceptual relevance.

Each study was coded based on full-text analysis and categorized by its alignment with the research questions. The mapping is shown in Table~\ref{tab:studies_per_rq}.

\begin{table}[h]
\footnotesize
\caption{Studies mapped to each Research Question}
\label{tab:studies_per_rq}
\begin{tabular}{|p{4.6cm}|p{3cm}|}
\hline
\textbf{Research Question} & \textbf{Studies} \\
\hline
\textbf{RQ1:} Socio-technical conditions and practices leading to smell emergence & 
S1\nocite{S1}, S2, S3, S4, S5, S6, S7, S8, S9, S10, S12, S13, S14, S15, S16, S17 \\
\hline
\textbf{RQ2:} Evolution and detection of code and community smells & 
S1, S3, S4, S5, S6, S8, S10, S11, S12, S13, S14, S15, S16, S18 \\
\hline
\textbf{RQ3:} Community smells as precursors or reinforcers of technical degradation & 
S1, S2, S3, S4, S5, S6, S7, S8, S9, S10, S11, S12, S13, S14, S15, S16, S17, S18 \\
\hline
\end{tabular}
\end{table}

\section{Findings from Formal Literature}
\label{sec:results_academic}

The analysis of the 18 selected peer-reviewed studies reveals not only isolated instances of socio-technical degradation but a web of interdependent conditions that sustain and reinforce smell dynamics in code samples. While the results are structured according to each research question (RQ), the cross-cutting connections among them, particularly the patterns of neglect, invisibility, and informal governance that jointly underpin the emergence, evolution, and entanglement of Code and Community Smells.

\subsection{RQ1 – Socio-technical conditions and practices leading to smell emergence}

\patterntitle{Fragmented communication and lack of structured review} The lack of structured communication and review mechanisms in the maintenance of code samples facilitates a dual-layered degradation: technical anomalies are introduced without contextual scrutiny, and social disconnection prevents their later identification or correction. Studies \textbf{S2}, \textbf{S4}, and \textbf{S6} show that asynchronous contributions, are rarely subject to peer validation. \textcolor{black}{This leads not only to unchecked quality issues but also to an erosion of collective memory, as illustrated by excerpts like: \enquote{several community examples are uploaded with minimal or no peer review…} [S2].} With this, developers are unable to trace the rationale behind structural decisions, and smells become normalized through reuse. This aligns with findings from \cite{Palomba2018, DeStefano2021} and complements observations by \cite{Soldani2021grey} on the invisibility of quality work. Notably, the absence of communication may not merely precede technical degradation but act as a social smell in itself, co-occurring with structural anomalies. Given the frequent replication of code samples, such degradation may propagate, leading to what could be framed as socio-technical smell contagion \cite{Wang2023}. \textcolor{black}{For instance, a developer copying a flawed code sample for a new tutorial inadvertently propagates its smells, infecting a new part of the ecosystem's knowledge base.} Therefore, fragmented communication does not only allow smells to survive—it creates the conditions for them to spread and multiply within and beyond their original scope.

\patterntitle{Centralized ownership and bottlenecks}  Evidence from \textbf{S3}, \textbf{S4}, and \textbf{S10} reveals that when code samples are predominantly maintained by a single developer or a small subset of contributors, opportunities for collaborative quality control are significantly reduced. \textcolor{black}{This is reinforced by practitioners who report that such centralization creates bottlenecks, with one stating, \enquote{When only one person knows how to maintain the examples, updates can take weeks, stalling community contributions} [D5].} Studies \textbf{S8} and \textbf{S12} reinforce that lack of distributed participation correlates with long-term stagnation, where smells persist not due to unawareness, but due to social dynamics that discourage intervention. This phenomenon aligns with concerns raised in open-source governance literature \cite{Annunziata2023, Lambiase2024}, which distinguish between healthy stewardship and exclusive gatekeeping. Over time, centralized ownership transforms code samples into socially protected but technically vulnerable artifacts—untouched not because they are robust, but because they are perceived as someone else’s responsibility.

\patterntitle{Devaluation of instructional artifacts} Despite the availability of static analysis tools such as Designite or SonarQube, studies \textbf{S1}, \textbf{S5}, \textbf{S13}, and \textbf{S14} reveal that these technologies are rarely applied to code samples. \textbf{S7} suggests that samples are commonly excluded from analysis pipelines due to their illustrative or instructional framing. This exclusion reflects a broader cultural dissonance: while tools exist to detect code anomalies, the artifacts in question are socially perceived as too trivial to merit scrutiny. This perception not only prevents refactoring but also strips samples of their pedagogical potential, transforming them into carriers of outdated or flawed design idioms. Moreover, the integration of tooling into collaborative workflows is rarely adapted to accommodate auxiliary modules, leaving a blind spot in quality assurance. The persistent underutilization of automated tools in these contexts is thus not a technical limitation, but a socio-cultural one.

\subsection{RQ2 – Evolution and detection of code and community smells}

Building on the emergence factors discussed above, this section examines how those same conditions shape the trajectory of smell persistence. The longitudinal dimension of quality assessment exposes another axis of exclusion: even when tools and practices exist to detect smells, they often fail to encompass code samples.

\patterntitle{Exclusion from longitudinal tracking} Despite advancements in tracking code quality over time, the longitudinal evolution of smells in code samples remains largely uncharted. Studies \textbf{S3}, \textbf{S6}, and \textbf{S14} employ snapshot-based approaches to map technical debt trajectories, yet they overwhelmingly target core modules, overlooking instructional artifacts. \textbf{S4} and \textbf{S16} confirm that auxiliary code is seldom included in quality assessments, resulting in a silent accumulation of long-lived smells. This omission stems from both technical and cultural factors: code samples often exist outside standard monitoring pipelines, lack change frequency, and are excluded from versioning metadata. \textbf{S18} advocates the use of contributor metadata to support evolutionary modeling, but such techniques face friction when applied to socially fragmented and historically shallow artifacts like samples. These gaps underscore a broader issue: existing quality models are structurally biased toward artifacts that conform to traditional development patterns.

\patterntitle{Onboarding gaps and turnover} Studies \textbf{S5}, \textbf{S11}, and \textbf{S15} demonstrate that contributor transitions—particularly the departure or reallocation of key maintainers—interrupt ongoing remediation cycles in code samples. When onboarding structures are absent or informal, newcomers lack the contextual knowledge to recognize smells, let alone address them. \textcolor{black}{Study [S11] provides evidence for this, noting that \enquote{in projects with high turnover and no documented sample guidelines, 60\% of smell-introducing commits for sample code came from developers with less than three months of tenure.}} These findings echo broader insights from software engineering research \cite{Mumtaz2022, Tandon2024}, which link role turnover to socio-technical misalignment. In the context of code samples, the impact is compounded: code samples are often seen as entry points for new contributors, yet their complexity and lack of ownership deter sustained engagement. Over time, unresolved smells become normalized, and instructional modules devolve into technical liabilities—under-maintained not due to irrelevance, but due to their exclusion from community routines of care.

\patterntitle{Underutilization of hybrid metrics} Although studies \textbf{S1}, \textbf{S8}, and \textbf{S12} propose hybrid metrics that integrate technical indicators with social signals (e.g., contributor churn), such approaches remain largely underutilized. \textbf{S14} emphasizes the importance of temporally and contextually aware indicators, yet mainstream tools persist in treating code quality as a purely structural concern. This structural bias overlooks the socio-organizational conditions under which smells emerge and persist. Moreover, while calls for hybrid monitoring abound \cite{GAROUSI2019101, Wang2023}, operational frameworks that translate these metrics into actionable insights are rare. The disjunction between detection capability and governance practice reflects a deeper issue: without integrating collaboration dynamics into quality strategies, instructional modules remain outside the scope of preventive or corrective action.

\subsection{RQ3 – Community smells as precursors or reinforcers of technical degradation}

Expanding on the previous RQs, this section delves into the reciprocal nature of socio-technical smells. Community Smells do not merely coincide with Code Smells—they actively enable, reinforce, and are sustained by them.

\patterntitle{Social fragmentation and structural flaws} Studies \textbf{S3}, \textbf{S4}, \textbf{S6}, and \textbf{S9} identify collaboration breakdowns—particularly \textit{Radio Silence} and \textit{Organizational Silos}—as consistent antecedents to complex code smells. These social fractures often render code samples analytically invisible: they are not discussed, not owned, and thus not maintained. \textbf{S14} and \textbf{S15} add that technical and social anomalies frequently co-occur, especially in modules with unclear accountability. This convergence reinforces recent work on smell entanglement \cite{Chen2024, Porta2024}, where dysfunction in coordination not only fails to prevent smells, but actively creates the organizational void in which they persist. Unlike core modules that benefit from structured ownership and review, code samples exist at the margins—where social fragmentation functions less as an incidental condition and more as a systemic enabler of technical debt.

\patterntitle{Governance voids and degradation loops} The absence of governance structures—such as defined roles or routine peer review—produces governance voids where smells not only emerge but persist through cycles of disengagement. Studies \textbf{S1}, \textbf{S10}, and \textbf{S13} show that in these contexts, contributors perceive code samples as high-cost, low-reward artifacts. \textbf{S7} and \textbf{S4} document how this perception reinforces neglect: smells discourage engagement, and disengagement reinforces smell persistence. Over time, this modules devolve into “dead zones” of maintenance, where structural anomalies go unobserved and unaddressed. These dynamics reflect the nature of socio-technical debt as not merely an accumulation of flaws, but as a systemic feedback loop of invisibility, dis-ownership, and abandonment \cite{Soldani2021grey, Baabad2020}.

\patterntitle{Degraded examples and community health} Studies \textbf{S2}, \textbf{S8}, \textbf{S12}, and \textbf{S17} highlight that technically flawed code samples lacking maintainer responsiveness are frequently avoided, abandoned, or rewritten. \textbf{S16} emphasizes that such artifacts, once deprecated, lose their instructional function, severing an important bridge for newcomers into the ecosystem. These degraded examples not only fail to teach—they actively repel. The lack of upkeep communicates exclusion, reducing contributor confidence and weakening the social scaffolding that underpins collaborative growth. Over time, this erodes the community’s cognitive continuity and onboarding efficacy. Rather than being inert failures, smell-laden samples operate as negative signals—markers of organizational entropy and entry barriers—especially when formal remediation routines are absent \cite{Menezes2019, Menezes2022}.

\section{Grey Literature Review (GLR)}
\label{sec:glr}

\subsection{Method and Source Selection}

To complement the findings from the academic literature and address contextual gaps regarding code samples in open-source software ecosystems, a Grey Literature Review (GLR) was conducted. This approach aimed to integrate practitioner insights on the emergence, evolution, and co-occurrence of \textit{Code Smells} and \textit{Community Smells} in open-source \textit{code samples}. In particular, the GLR focused on scenarios involving decentralized maintenance, informal collaboration, and tooling limitations—conditions often absent from formal publications.

The review followed multivocal literature guidelines~\cite{GAROUSI2019101}, incorporating both academic and non-academic sources. These guidelines emphasize the relevance of grey sources in domains where industrial experience precedes theoretical consolidation, and where informal practices shape software quality. In the context of OSSECOs, and especially in artifacts such as code samples, technical debt is frequently discussed in blog posts, community reports, and development forums rather than peer-reviewed channels.

\textbf{Search Strategy.} The search was conducted using Google, following the "effort-bounded" strategy described in~\cite{GAROUSI2019101}. \textcolor{black}{This strategy involved screening the first 10 pages of results for each query and stopping when two consecutive pages yielded no new relevant sources, indicating a point of theoretical saturation for the scope of this review.} Queries combined terms related to software quality, code samples, and socio-technical smells, such as \texttt{"code smell"}, \texttt{"community smell"}, \texttt{"developer communication"}, \texttt{"refactoring"}, and \texttt{"open source"}.

Results were constrained to publicly accessible content written in English or Portuguese, published between 2013 and 2024. Priority was given to sources that demonstrated practitioner credibility (e.g., named authors, affiliations) and direct relevance to the research questions.
\textbf{Source Types.} The review considered four main categories:
\begin{itemize}
    \item Technical blogs (e.g., Dev.to, Medium);
    \item Community discussions (e.g., Reddit, Stack Overflow threads);
    \item Reports and datasets in community repositories (e.g., Zenodo, GitHub releases);
    \item Informal white papers or preprints authored by industry professionals.
\end{itemize}
Platforms such as Dev.to and Reddit were prioritized due to open access, high engagement, and traceable authorship. Sources with paywalls or anonymous/unverifiable content were excluded.

\begin{figure}[h]
\centering
\includegraphics[width=\linewidth]{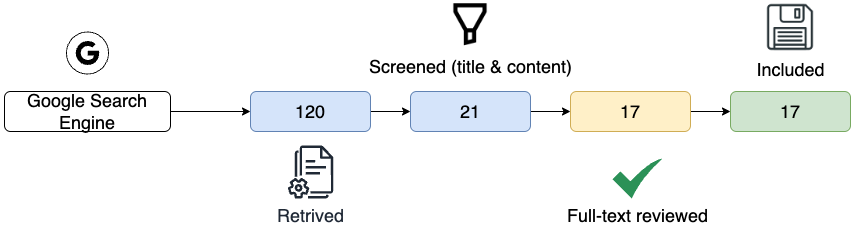}
\caption{Selection process for the Grey Literature Review (GLR)}
\label{fig:glr_flow}
\end{figure}

\textbf{Inclusion and Filtering.} An initial corpus of 120 documents was retrieved. The screening process followed three stages as show on Figure~\ref{fig:glr_flow}, and final corpus consisted of 17 documents.

\begin{enumerate}
    \item \textit{Preliminary scan}: Titles and samples were analyzed for thematic alignment;
    \item \textit{Full-text review}: Each candidate was read to assess clarity, empirical grounding (e.g., examples, metrics, tools), and relevance to the RQs;
    \item \textit{Validation}: Only documents with observable socio-technical content and alignment with code sample contexts were retained.
\end{enumerate}

\subsection{Corpus Overview}

Seventeen practitioner-oriented documents were selected and analyzed D1-D17~\cite{D1,D2,D3,D4,D5,D6,D7,D8,D9,D10,D11,D12,D13,D14,D15,D16,D17}. Each source was thematically coded and classified according to its type, relevance, and alignment with the guiding research questions. Table~\ref{tab:glr_docs_per_rq} summarizes the mapping of documents to the research questions.

\begin{table}[htbp]
\footnotesize
\caption{Grey literature documents mapped to each Research Question}
\label{tab:glr_docs_per_rq}
\centering
\begin{tabular}{|p{4.6cm}|p{3cm}|}
\hline
\textbf{Research Question} & \textbf{Documents} \\
\hline
\textbf{RQ1:} What socio-technical conditions and practices lead to the emergence of Code Smells and Community Smells in code samples? & 
D1, D3, D4, D5, D6, D8, D10, D14, D16, D17 \\
\hline
\textbf{RQ2:} How do Code Smells and Community Smells evolve over time in open-source code samples, and how can this evolution be detected? & 
D1, D2, D3, D6, D8, D9, D11, D13, D15, D16, D17 \\
\hline
\textbf{RQ3:} How do Community Smells signal or reinforce the emergence and persistence of Code Smells in code samples? & 
D2, D4, D5, D7, D10, D11, D12, D13, D14, D15, D16, D17 \\
\hline
\end{tabular}
\end{table}

\section{Findings from Grey Literature}
\label{sec:results_glr}

The review of 17 practitioner-oriented documents revealed a rich but fragmented landscape of socio-technical concerns surrounding the emergence, persistence, and co-occurrence of \textit{Code Smells} and \textit{Community Smells} in open-source code samples. These sources—blogs, white papers, discussion threads, and technical evaluations—offered grounded insights into the practical realities faced by developers. Unlike the structured rigor of formal research, the grey literature surfaces experiential knowledge, tool-based reflections, and implicit observations of governance failures. Although explicit references to Community Smells were rare, many documents captured social decay patterns congruent with those described in academic work.

\subsection{RQ1 – Socio-technical conditions and practices leading to smell emergence}

\patterntitle{Ownership ambiguity and lack of stewardship}Sources such as \textbf{D1}, \textbf{D5}, and \textbf{D8} expose a recurring pattern in open-source repositories: code samples are frequently contributed without any mechanism for long-term stewardship. This absence of clear ownership generates ambiguity regarding who holds responsibility for maintaining, reviewing, or improving these artifacts. In many cases, even when flaws are identified, contributors hesitate to intervene, not out of lack of competence, but due to uncertainty over authority and accountability boundaries. As illustrated in \textbf{D14}, a widely reused snippet with known flaws remained unaltered for several iterations simply because, in the words of a contributor, \textcolor{black}{\enquote{no one owns the fix.}}

This form of socio-technical inertia exemplifies a condition that might be described as a “latent decay zone”: an artifact whose deterioration is not due to active neglect, but to the passive absence of governance. Over time, such zones normalize quality erosion. The samples, become central to documentation and onboarding, embedding smells into the developer experience and creating invisible technical debt. These observations align with findings in formal literature regarding informal governance structures and their impact on diferent modules \cite{Annunziata2023, Lambiase2024}, but the practitioner narratives here emphasize the emotional and cognitive cost of contributing to ambiguous artifacts—where taking initiative feels risky and unrewarded.

\patterntitle{Reactive and undervalued quality control} Documents such as \textbf{D3}, \textbf{D4}, and \textbf{D7} describe a recurrent dynamic in the treatment of instructional code: quality control is neither preventive nor systematic, but instead activated only after a visible failure occurs. Refactoring efforts are framed as reactive responses to bugs, user confusion, or documentation drift—rather than as part of a proactive maintenance culture. This temporal deferral does not indicate a lack of awareness, but rather reflects a triage-based prioritization in which production-facing code absorbs most of the available attention and effort. As noted in \textbf{D11}, one contributor remarks: \textcolor{black}{\enquote{We’ll clean it up next time we teach this,}} suggesting that instructional artifacts are implicitly considered transient or disposable—even when reused across multiple contexts.

This perception undermines their inclusion in formal quality assurance routines. The result is a two-tiered system in which code samples are excluded from the normative expectations of code hygiene. Such asymmetries perpetuate technical anomalies not by accident, but by design—implicitly codifying a division between what deserves care and what can be “fixed later.”

\patterntitle{Tool distrust and workflow misalignment} Practitioner accounts from \textbf{D9} and \textbf{D13} reflect a pattern of skepticism toward the applicability and reliability of automated quality assessment tools in the context of instructional or auxiliary code. Tools such as SonarQube and CodeScene are often described as producing excessive false positives or lacking the contextual awareness necessary to assess examples fairly. \textcolor{black}{This sentiment was echoed in a community thread, where a developer mentioned, \enquote{We had to disable Sonar for the /examples folder. The noise was overwhelming, and every report felt like a false positive in our context} [D9].} As a result, contributors frequently exclude example files from analysis pipelines or disable specific rule checks. These practices, while pragmatic, reflect a deeper misalignment between tooling assumptions and the informal nature of code samples. This dynamic mirrors observations in the formal literature, where the mere presence of detection tools does not translate into systematic use—particularly in artifacts considered marginal or temporary.

\patterntitle{Neglect of pedagogical intent} Several practitioner-oriented sources implicitly reveal a divergence between the initial instructional purpose of code samples and the absence of mechanisms to preserve that purpose over time. Documents \textbf{D6} and \textbf{D15}, for instance, describe scenarios in which tutorials or technical blogs are published with illustrative code that aligns closely with a given toolset or framework version. However, after publication, these examples are seldom maintained. As dependencies evolve, syntactic conventions shift, or APIs deprecate, the code remains static—causing it to become outdated or even misleading to new learners. This form of degradation transcends simple technical obsolescence. It undermines the artifact’s pedagogical credibility, transforming what was once a didactic resource into a potential source of confusion or error. The issue is not that the code ceases to compile, but that it ceases to teach effectively.

\subsection{RQ2 – Evolution and detection of code and community smells}

\patterntitle{Exclusion from longitudinal quality strategies}  
Several practitioner sources indicate that while software quality is increasingly monitored through temporal metrics—such as commit frequency, churn, and defect density—these efforts are overwhelmingly focused on core production code. For instance, \textbf{D3} discusses how quality metrics are tracked in enterprise projects, but makes no mention of auxiliary artifacts like code samples. The dataset in \textbf{D2} further reinforces this exclusion: although aimed at supporting smell detection at scale, it omits code examples due to their inconsistent structure and lack of traceable authorship. This omission is not merely a matter of pipeline configuration; it reveals a systemic misclassification of code samples as static or disposable. In \textbf{D13}, CodeScene is critiqued for failing to contextualize small-scale scripts, contributing to what one user terms a \textcolor{black}{\enquote{blind spot in smell surveillance.}} Consequently, code samples become analytically invisible, despite being versioned and circulated across repositories.

\patterntitle{Recognized temporal drift without diagnosis}  
In contrast to this analytical invisibility, developers consistently report informal recognition of smell accumulation in code samples—often after these artifacts cause misunderstanding or friction. In \textbf{D11}, an illustrative tutorial was only flagged as problematic after users misapplied it, prompting a late-stage revision. While tools exist for historical smell analysis \cite{D17}, they are seldom applied to code samples due to a lack of structured ownership, tagging, or annotated histories. Even in projects with version control, as observed in \textbf{D15}, developers rarely document the rationale behind updates to examples, making it difficult to reconstruct their semantic evolution. This absence of longitudinal traceability leads to what \textbf{D1} frames as “temporal opacity”: developers know that decay is occurring but cannot localize its onset or causes. As a result, maintenance becomes reactive.

\subsection{RQ3 – Community smells as precursors or reinforcers of technical degradation}

\patterntitle{Social abandonment and maintenance voids} The practitioner sources (\textbf{D4}, \textbf{D5}, \textbf{D10}) describe the progressive neglect of auxiliary code—particularly instructional examples—as project priorities shift. Over time, these artifacts lose alignment with the main system, yet remain in the repository. \textbf{D8} illustrates that even small contributions aimed at correcting outdated examples often receive no feedback, creating a perception that such modules are no longer under community care. This form of neglect is not merely technical but social: it signals that these files fall outside any active stewardship. The concept of "maintenance voids" refers to zones in the codebase where no individual or group assumes responsibility for upkeep. The invisibility is often unintentional, but the absence of interaction, ownership, or responsiveness turns these artifacts into dormant components that accumulate technical debt.

\patterntitle{Unclear roles and mentorship erosion} Practitioner accounts from \textbf{D14} and \textbf{D16} reveal that contributors often hesitate to engage with instructional modules due to unclear social protocols. In these contexts, developers express uncertainty regarding review expectations, contribution boundaries, and ownership. Questions such as “who maintains this?” or “what constitutes a valid change here?” remain unanswered, leading to inaction. This hesitation illustrates a deeper issue: the absence of mentorship structures and role transparency. When contributors cannot discern what is expected of them, psychological friction increases. These dynamics correspond with recognized community smells, particularly \textit{Weak Onboarding} and \textit{Unclear Roles}, which hinder integration and suppress initiative.

\patterntitle{Low-quality samples as negative signals} Documents \textbf{D15} and \textbf{D17} illustrate how technically degraded instructional examples convey more than technical shortcomings. For prospective contributors, these artifacts function as proxies for project health and governance culture. Rather than serving as onboarding tools, such examples deter engagement by signaling a disconnect between community ideals and practical maintenance routines. This phenomenon reflects a negative social signal: a structural artifact that communicates unintentional messages about community priorities. \textcolor{black}{A developer on a forum [D15] stated this clearly: \enquote{When I see a project with sloppy, outdated examples, I immediately question the quality of the core product and the attentiveness of the maintainers. It’s a huge red flag.}} When example files appear neglected, the implied message is one of disengagement. This perception undermines trust not only in the module itself, but in the broader development process.

\patterntitle{Perceived disorganization} Documents \textbf{D11} and \textbf{D12} describe situations where small inconsistencies in code samples—such as outdated library references or naming mismatches—accumulate and foster a broader sense of project disorganization. These technical inconsistencies may appear trivial in isolation, yet function as collective red flags suggesting a lack of coordination or oversight. This perception of disarray undermines what can be termed collaborative perception—the belief that a given artifact is maintained through a coherent community process. When examples appear inconsistent, contributors may infer that broader collaboration practices are equally fragmented, reducing their willingness to engage.

\section{Threats to Validity}
\label{sec:threats}

This review is subject to known limitations associated with multivocal studies. Potential threats are: \textbf{\textit{Search and Retrieval}} - In the academic phase, three major digital libraries were used. Still, relevant studies may have been excluded due to terminology variation. In the grey literature phase, results were influenced by Google’s ranking, which may limit source diversity. To address this, effort-bounded strategies were used and searches were conducted in English and Portuguese, across multiple iterations. \textbf{\textit{Selection and Inclusion}} - The study followed predefined inclusion and exclusion criteria across both literature types. For the grey literature, source credibility was assessed based on traceable authorship and thematic relevance. Nonetheless, grey sources often lack consistent structure, and some judgment calls were needed during screening and coding. \textbf{\textit{Synthesis and Interpretation}} -Thematic synthesis involves interpretation, which introduces the risk of bias. To mitigate this, coding was conducted independently for formal and grey sources, and only integrated in the final synthesis stage. Practitioner quotes were paraphrased to preserve anonymity and analytical consistency. \textbf{\textit{Transferability}} - Findings are drawn from open-source environments and may not generalize to closed or enterprise settings. Additionally, grey sources represent only publicly shared practices. However, the convergence of findings across literature types suggests that the identified patterns may reflect broader socio-technical dynamics.

\section{Discussion and Implications}
\label{sec:discussion}

\textcolor{black}{This section consolidates and interprets the findings from both literature streams. First, we present a summary of the nine identified patterns that serve as the foundation for our analysis (Table~\ref{tab:pattern_summary}). Following this, we synthesize these patterns across the research questions, connecting them to broader socio-technical theories. Finally, we derive concrete implications for key stakeholders within OSSECOs.}

\begin{table*}[htbp]
\caption{Synthesis of the Nine Socio-Technical Patterns Identified}
\label{tab:pattern_summary}
\small
\begin{tabularx}{\textwidth}{|l|X|l|}
\hline
\textbf{Pattern Name} & \textbf{Description} & \textbf{Primary Evidence} \\
\hline
\multicolumn{3}{|c|}{\textbf{RQ1: Emergence Conditions}} \\
\hline
\textbf{1. Fragmented Communication} & Lack of structured review enables unverified contributions, which introduce and normalize smells. & S2, S4, S6, D14 \\
\hline
\textbf{2. Centralized Ownership} & A single developer or a small group acts as a bottleneck, reducing collaborative quality control and causing technical stagnation. & S3, S4, S10, D1, D5 \\
\hline
\textbf{3. Tooling Exclusion \& Distrust} & Instructional artifacts are culturally devalued and excluded from static analysis pipelines, often due to perceived noise or false positives. & S1, S7, D9, D13 \\
\hline
\multicolumn{3}{|c|}{\textbf{RQ2: Evolution and Persistence}} \\
\hline
\textbf{4. Longitudinal Invisibility} & Code samples are omitted from long-term quality tracking, which allows smells to accumulate silently over time. & S4, S16, D2, D3 \\
\hline
\textbf{5. Onboarding Gaps} & High contributor turnover, combined with a lack of mentorship, obstructs remediation cycles as newcomers lack contextual knowledge. & S5, S11, D14, D16 \\
\hline
\textbf{6. Reactive Refactoring} & Quality control is reactive and activated only after a visible failure, rather than being a proactive maintenance practice. & S7, S1, D3, D4, D7 \\
\hline
\multicolumn{3}{|c|}{\textbf{RQ3: Co-occurrence and Reinforcement}} \\
\hline
\textbf{7. Social Abandonment} & A lack of stewardship turns examples into "maintenance voids," where social inaction leads to recursive technical degradation. & S3, S9, D8, D10 \\
\hline
\textbf{8. Governance Voids} & The absence of defined roles and routines creates feedback loops where flawed samples discourage engagement, which in turn reinforces decay. & S1, S10, D14, D16 \\
\hline
\textbf{9. Negative Social Signaling} & Degraded examples act as negative proxies for project health, eroding contributor trust and weakening community onboarding. & S17, D15, D17 \\
\hline
\end{tabularx}
\end{table*}

\subsection{Cross-RQ Synthesis}

The integrated synthesis of formal and grey literature reveals that the emergence, persistence, and co-occurrence of \textit{Code Smells} and \textit{Community Smells} in open-source code samples are not isolated events. Rather, they represent the entangled outcome of recurring socio-technical conditions that span the three research questions.

\textbf{In relation to RQ1}, the findings indicate that \textbf{governance gaps}—including undefined ownership, informal review practices, and ad hoc quality routines—are key enablers of smell emergence. Formal studies such as \textbf{S1}, \textbf{S3}, and \textbf{S6} show that code samples under centralized or ambiguous stewardship are less likely to receive peer validation or undergo refactoring. Grey literature sources (\textbf{D1}, \textbf{D4}, \textbf{D5}) reinforce this by highlighting contributor perceptions that these artifacts are “unclaimed,” which inhibits intervention and promotes decay. Governance, in this context, functions not merely as an organizational layer, but as a structural mediator of both technical and social degradation.

\textbf{Regarding RQ2}, the review identifies a persistent \textbf{visibility paradox}. Although code samples are widely reused and disseminated (\textbf{S9}, \textbf{D6}), they are often excluded from quality assurance pipelines (\textbf{S7}, \textbf{D13}) and smell detection mechanisms (\textbf{D9}, \textbf{D13}). Tools are available, but their assumptions and configurations are misaligned with the nature of instructional code. Consequently, smells in these artifacts persist not due to the absence of detection capability, but due to a lack of contextual trust and routine adoption. This gap, as noted in \textbf{S14} and \textbf{D3}, results in "invisible debt zones" where quality decay remains analytically hidden.

\textbf{In response to RQ3}, the co-evolution of code and community smells emerges as a systemic rather than incidental phenomenon. Empirical evidence (\textbf{S2}, \textbf{S4}, \textbf{S5}) shows that social fragmentation—such as \textit{radio silence}, role ambiguity, and siloed communication—often precedes technical decay. These findings are echoed in grey sources (\textbf{D10}, \textbf{D14}, \textbf{D16}), where contributors describe scenarios of social abandonment that leave instructional artifacts without oversight. Such conditions reinforce the notion of \textbf{socio-technical incongruence}, in which organizational structures are misaligned with the artifact’s intended lifecycle. \textcolor{black}{This phenomenon can be seen as a manifestation of Conway's Law, where the fragmented communication structure of the community is mirrored in the fragmented and decaying quality of its artifacts.}

Finally, across all RQs, the findings converge on the existence of \textbf{feedback loops of degradation}. Initial neglect—driven by centralized ownership (\textbf{S3}, \textbf{D8}) or misaligned processes (\textbf{S6}, \textbf{D7})—leads to contributor disengagement (\textbf{S17}, \textbf{D14}), which in turn perpetuates invisibility and decay. Grey literature further enriches this picture by exposing subjective and symbolic dimensions, such as hesitancy to intervene (\textbf{D14}), absence of mentorship (\textbf{D16}), and negative signaling from degraded examples (\textbf{D15}, \textbf{D17}). These recursive dynamics are rarely addressed by current models of quality assurance, suggesting that smells in code samples are not merely technical artifacts—but indicators of deeper structural and cultural misalignment in open-source ecosystems.

\textcolor{black}{Notably, the multivocal approach was critical in uncovering these dynamics. While formal studies (e.g., S1, S3, S6) identified structural issues like \textit{governance gaps}, the grey literature (e.g., D1, D9, D14) provided the crucial context behind them: contributor hesitancy, tool distrust, and the perception of artifacts as “unclaimed.” This synthesis demonstrates that a traditional SLR would have identified the symptoms, whereas the MLR revealed the underlying socio-technical drivers, thus providing a more holistic and actionable understanding.}

\subsection{Implications for Practitioners, Managers, and Researchers}
\label{sec:implications}

\textcolor{black}{The results highlight that code samples, play a strategic role in open-source ecosystems. Their neglect reflects not only technical debt but also organizational fragility. Treating code samples as a first-class citizen is essential to support onboarding, sustain community health, and ensure long-term maintainability.}

\textcolor{black}{\textit{For managers and maintainers}, our findings provide actionable guidelines for targeted governance. We recommend that they: (i) assign explicit stewards to ensure accountability (e.g., via \texttt{CODEOWNERS} files); (ii) conduct periodic reviews for both technical correctness and instructional quality; (iii) establish lightweight verification pipelines (e.g., adapted static analysis, build validation); and (iv) maintain structured contribution workflows to manage community feedback and preserve quality. These actions directly translate our identified patterns into practice.}

\textcolor{black}{\textit{For practitioners and developers}, the nine patterns serve as a diagnostic framework. For instance, to counter "Reactive Refactoring," a developer can configure a bot like Dependabot for automated updates. To mitigate "Exclusion from Tooling", a developer can create a dedicated, lenient configuration profile for static analysis tools, ensuring meaningful quality assurance without the noise of false positives.}

\textcolor{black}{\textit{For tool developers}, standard quality tools are miscalibrated for code samples, which prioritize instructional clarity over production-level complexity. This creates excessive false positives, causing maintainers to ignore the results. This limitation exposes a measurement gap, as current tools are not calibrated for instructional artifacts~\cite{8870139}. Incorporating social indicators—such as ownership gaps or contributor turnover—can improve detection in these low-visibility artifacts.}

\section{Conclusion and Future Work}
\label{sec:conclusion}

This multivocal review analyzed how \textit{Code Smells} and \textit{Community Smells} emerge, persist, and intersect in code samples within open-source software ecosystems. Based on 18 peer-reviewed studies and 17 practitioner-oriented documents, the synthesis identified socio-technical patterns that sustain degradation in these code samples.

Despite their small size and instructional purpose, code samples exhibit recurrent neglect due to centralized ownership, informal governance, and absence from structured quality assurance workflows. These conditions contribute to sustained smell accumulation and signal deeper coordination and stewardship gaps. In several cases, technical decay aligned with organizational fragility, reflecting a lack of role clarity and breakdowns in onboarding routines.

The evidence challenges the notion that code samples are inconsequential to ecosystem health. Instead, their condition often anticipates broader risks to sustainability, trust, and contributor engagement. Smells in code samples are rarely isolated—they indicate latent structural and social tensions.

Future research should aim to operationalize the identified patterns into hybrid indicators that combine structural metrics (e.g., smell lifespan, ownership dispersion) with behavioral signals (e.g., contributor churn, review latency). Empirical validation across diverse repositories can assess generalizability, while the design of lightweight interventions—such as peer review routines tailored to auxiliary code—may support early detection and mitigation.

\textcolor{black}{Finally, the verification of authors in the grey literature was a quality control step to ensure source credibility, as per MLR guidelines~\cite{GAROUSI2019101}. An analysis of differing author perceptions was outside our study's scope. However, we agree this is a valuable research direction. Understanding these perceptions would require direct methods, like surveys or interviews. We have added this as a promising avenue for future work.}

\textcolor{black}{
\section*{Data Availability}
\label{sec:data}
}

\textcolor{black}{The supplementary artifact containing the full mapping of sources, textual evidence, and thematic codes is openly available on Figshare at: \url{https://doi.org/10.6084/m9.figshare.29441441.v1}.}

\begin{acks}
This work was funded by the Federal University of Mato Grosso do Sul and by Brazilian funding agencies CAPES (Finance Code 001), FAEPEX (grants: 3404/23 and 2382/24) and the Brazilian funding agency FUNDECT (Call No. 42/2024).
\end{acks}


\end{document}